\documentstyle{article}
\begin{document}
\begin{titlepage}
\bigskip

YCTP-P39-91  October 24 1991

\bigskip
\centerline{GEOMETRICAL LATTICE MODELS FOR  $N=2$ SUPERSYMMETRIC THEORIES}
\centerline{IN TWO DIMENSIONS}
\vspace{2cm}
\centerline{H.Saleur\footnote{On leave from Spht Cen Saclay 9119 Gif Sur Yvette
Cedex France}\footnote{Work supported in part by DOE grant DE-AC0276ER03075 and
by a Packard Fellowship for Science and Engineering}}
\centerline{Physics Department}
\centerline{Yale University}
\centerline{New Haven CT06511}
\centerline{USA}
\vspace{2cm}
\centerline{Abstract}
\bigskip
We introduce in this paper two dimensional lattice models
 whose continuum limit
belongs to the $N=2$ series. The first kind of model is integrable and
obtained through a
geometrical reformulation, generalizing results known in the $k=1$ case,  of
the $\Gamma_{k}$ vertex models (based on the quantum algebra $U_{q}sl(2)$ and
representation of spin $j=k/2$). We demonstrate in particular that at the $N=2$
point, the free energy of the $\Gamma_{k}$ vertex  model can be obtained
exactly
by counting arguments, without any Bethe ansatz computation, and we
exhibit lattice operators that reproduce the chiral ring. The
second class of models is more adequately described in the language of twisted
$N=2$ supersymmetry, and consists of an infinite series of multicritical
polymer

points, which should lead to experimental realizations.
 The presence of $N=2$ in that case is traced back to the Parisi Sourlas
supersymmetry of the lagrangians usually used to replace $n\rightarrow 0$
limits. Boundary conditions as well as fermionic and bosonic variables are
geometrically interpreted. Moreover it turns out
that  the exponents $\nu=(k+2)/2(k+1)$ for
these multicritical  polymer points coincide with the old phenomenological
formulas of
Flory. We therefore confirm that these formulas are {\bf exact}
 in two dimensions, and
suggest that their unexpected validity is due to non renormalization theorems
 for
the $N=2$ underlying theories. We also discuss the status of the much discussed
theta point for polymers in the light of $N=2$ renormalization group flows.
\end{titlepage}
\section{Introduction}

$N=2$ theories are one of the most interesting example of conformal theories in
two dimensions. A large subset of them  possess in particular a very useful
 Landau Ginzburg
 description \cite{KMS89,VW89,M89}
, whose efficiency is ensured by non renormalization theorems\cite{HW}
 \footnote{The $N=0$ and
$N=1$ minimal series also possess in principle Landau Ginzburg descriptions,
however these cannot really be used to obtain precise information on the
system, like critical exponents.}. The Landau Ginzburg description makes also
transparent the connection with singularity theory \cite{VW89,M89},
 and explains nicely the
origin of $ADE$ type classification.

Unfortunately these beautiful structures have not so far been much observed. It
is known in principle how some special points of the $\Gamma_{k}$ vertex
 models\cite{DFSZ88}, the XY
(Gaussian) model or the Ashkin Teller \cite{FS,GR87,YZ87} model possess $N=2$
supersymmetry. However it has  not been clear so far what is special
 about these
points, and therefore how to tune the parameters to observe them. Also one
would
like the Landau Ginzburg description, and the chiral ring, to play some
more illuminating  role in these identifications, and make sense physically, as
is partly the case for $N=0$ theories \cite{Z86}.

In a preceding paper \cite{S91} we addressed these questions in the simplest
cas
   e
$k=1$. We showed that the best point of view was to consider twisted $N=2$
\cite{W88,EY90}, and that realizations of such theories
 were nicely provided by the
geometrical problems of polymers and percolation. $N=2$ supersymmetry was
manifest in the structure of the correlators and the geometrical operators
algebra. The presence of supersymmetry was traced back to the Parisi Sourlas
\cite{PS80}
supersymmetry  of the lagrangians (involving bosons and fermions) that are used
in the description of geometrical models ($n\rightarrow 0$ limits). The
knowledge of supersymmetry allowed us also to predict some new exponents, like
the fractal dimension of the backbone of percolation in two dimensions.

The purpose of the present paper is to extend these observations to the case of
arbitrary $k$. We find in particular lattice analogs of the chiral ring in
integrable geometrical models, and
families of multicritical polymer points connected by the $N=2$ flow.

The organization is as follows. In the second section we work out the
reexpression of $N=2$ partition functions in terms of generalized Coulomb gas
that was started in \cite{DFSZ88}. In section three we study the $\Gamma_{k}$
vertex models using a new geometrical formulation that generalizes the $k=1$
case. This formulation involves bound states of strands carried by the edges of
the square lattice. It exhibits very clearly the symmetry properties of the
$q=exp(i\pi/k+2)$ points, at which the continuum limit is the $A_{k+1}$ modular
invariant for the $N=2$ series. For instance the free energy can be obtained in
a simple way at that point, without using any Bethe ansatz computation.
The continuum limit is indeed derived. We also
identify a set of lattice operators that reproduce the chiral ring in an
appropriate continuum limit. In the fourth section, we consider the duals (in
the sense of Coulomb gas in two dimensions) of
these models, and argue that they correspond to multicritical polymers.
Generalizing the analysis of \cite{S91}, the various sectors and the role of
fermions and bosons, are geometrically interpreted. We discuss the coincidence
of the exponents formulas with the so called Flory formulas \cite{F71}.

\section{Generalized Coulomb Gas
 Expressions for the $N=2$ Partition Functions}

The generalized lattice Coulomb gas was introduced in \cite{DFSZ88} to describe
the continuum limit of $SU(2)$ minimal coset models and of $\Gamma_{k}$ vertex
models based on spin $j=k/2$ representations of $U_{q}sl(2)$.
It involves as expected
\cite{R88} a free bosonic field and $Z_{k}$ parafermions, which are coupled
through boundary conditions. Introduce the partition function of a free bosonic
field with coupling $g$ and windings $2m\pi,2m'\pi$ along the generators
$\omega_{1},\omega_{2}$ of the torus, with $\tau=\omega_{2}/\omega_{1}$
\begin{equation}
Z_{mm'}(g)=\sqrt{\frac{g}{Im\tau}}\frac{1}{\eta\overline{\eta}}exp\left(
-\frac{\pi g}{Im\tau}\left|m-m'\tau\right|^{2}\right)
\end{equation}
where $\eta$ is Dedekind's function. Introduce also for a simply laced Lie
algebra $G$ with Coxeter number $k+1$ the partition functions $Z^{G}_{k}(r,s)$
of the $Z_{k}$ model of \cite{ZF85,GQ87} with boundary conditions
 twisted by $e^{2i\pi
r/k},e^{2i\pi s/k}$ (we will not need their exact expression in the following).
The generalized Coulomb partition function was then defined in \cite{DFSZ88}
\begin{equation}
Z^{G}_{gc}\left(g\right)=\sum_{r,s=0,\ldots,k-1}Z^{G}_{k}\left(r,s\right)\sum_{
\begin{array}{c}
m=r\mbox{ mod }k\\
m'=s\mbox{ mod }k
\end{array}}Z_{mm'}\left(g\right)\label{eq:zcoul}
\end{equation}
with central charge
\begin{equation}
c=\frac{3k}{k+2}\label{eq:c}
\end{equation}
In the following we shall concentrate on Lie algebras of $A$ type, and
henceforth suppress the superscript $G$. The coulomb partition functions
satisfy the duality relation
\begin{equation}
Z_{gc}\left(g\right)=Z_{gc}\left(\frac{1}{k^{2}g}\right)
\end{equation}

It was then shown in \cite{DFSZ88} that the $N=2$ partition functions for the
minimal $A_{k+1}$ series with central charge  (\ref{eq:c}) obtained
in \cite{RY87} could be rewritten as follows\footnote{The labels A
 (antiperiodic)
and P (periodic) refer to the fermion boundary conditions on the torus}
\begin{eqnarray}
Z'_{N=2,k}&=&\frac{1}{2}\left(Z_{AP}+Z_{PA}+Z_{AA}\right)\nonumber\\
&=&\sum_{r,s=0,\ldots,k-1}Z_{k}(r,s)\sum_{\begin{array}{c}
m=r\mbox{ mod }k\\
m'=s\mbox{ mod }k
\end{array}}\left(\frac{1}{2}+
\delta_{m\wedge m'\mbox{ mod }2}\right)Z_{mm'}(\frac{k+2}{2k})\label{eq:N=2}
\end{eqnarray}
The doubly periodic contribution is also known \cite{KMS89}to be
\begin{equation}
Z_{PP}=Tr(-)^{F}=k+1
\end{equation}
Notice that this index does not vanish.

We can now simplify the  expression (\ref{eq:N=2}) using the following
arguments. It was also shown in \cite{DFSZ88} how the generalized Coulomb gas
can be used to reexpress the partition functions of minimal models
\begin{equation}
\frac{SU(2)_{k}\otimes SU(2)_{l-2}}{SU(2)_{k+l-2}}
\end{equation}
with central charges
\begin{equation}
c=\frac{3k}{k+2}-\frac{6k}{l(l+k)}
\end{equation}
This reexpression involves the generalized Coulomb
gas with coupling
\begin{equation}
g=\frac{1}{k}-\frac{1}{l+k}
\end{equation}
plus some interactions between the winding numbers.
Consider now the particular case $l=2$ that is associated with lattice models
of
type $A_{k+1}$. For such models $c=0$ and one expects the partition function to
be a constant, equal to $k+1$ (the number of possible heights per lattice
site). On the other hand, following \cite{DFSZ88} one gets
\begin{equation}
Z_{A_{k+1}}=\sum_{r,s=0,\ldots,k-1}Z_{k}(r,s)\sum_{\begin{array}{c}
m=r\mbox{ mod }k\\
m'=s\mbox{ mod }k
\end{array}}Z_{mm'}\left(\frac{2}{k(k+2)}\right)\sum_{n=1}^{k+1}cos\left(
2\pi\frac{n}{k+2}m\wedge m'\right)
\end{equation}
The sum over exponents produces a $(k+2)\delta_{m\wedge m'\mbox{ mod }k+2}-1$.
Analysis of each of the two terms and use of the duality symmetry
gives therefore the formula
\begin{equation}
Z_{A_{k+1}}=k+1=\sum_{r,s=0,\ldots,k-1}Z_{k}(r,s)\sum_{\begin{array}{c}
m=r\mbox{ mod }k\\
m'=s\mbox{ mod }k
\end{array}}\left(2
\delta_{m\wedge m'\mbox{ mod }2}-1\right)Z_{mm'}\left(\frac{k+2}{2k}
\right)\label{eq:ramond}
\end{equation}
Notice that this can be reformulated as
\begin{equation}
k+1=\sum_{r,s=0,\ldots,k-1}Z_{k}(r,s)\sum_{\begin{array}{c}
m=r\mbox{ mod }k\\
m'=s\mbox{ mod }k
\end{array}}(-)^{m\wedge m'}Z_{mm'}\left(\frac{k+2}{2k}\right)
\end{equation}
This identity generalizes the Euler's identities. One deduces
 from (\ref{eq:N=2})
and (\ref{eq:ramond})
\begin{equation}
Z'_{N=2,k}=Z_{gc}\left(\frac{k+2}{2k}\right)+\frac{k+1}{2}=Z_{gc}\left(
\frac{2}{k(k+2)}\right)+\frac{k+1}{2}
\end{equation}
The full $N=2$ partition functions with projection on odd fermion number writes
therefore
\begin{equation}
Z_{N=2,k}=Z_{gc}\left(\frac{k+2}{2k}\right)=
Z_{gc}\left(\frac{2}{k(k+2)}\right)\label{eq:z}
\end{equation}
This identity was  originally derived in \cite{DFSZ} and independently in
 \cite{Y88}. It provides a first clue concerning the lattice models whose
continuum limit is $N=2$ supersymmetric. Indeed the generalized Coulomb
partition functions are the continuum partition functions for the $\Gamma_{k}$
vertex models based on the quantum algebra $U_{q}sl(2)$
and spin $j=k/2$.
 However this identification leaves totally obscure the origin of
the $N=2$ supersymmetry. At first sight, nothing distinguishes the
 $\Gamma_{k}$ vertex model at the value $q=exp(i\pi/k+2)$
from any other
point on its critical line \cite{DFSZ88}. The next section is devoted to a
geome
   trical
formulation that will clarify this issue.

\section{Geometrical Reformulation of the $\Gamma_{k}$ Vertex Models}

\subsection{Generalities}

The weights of the $\Gamma_{k}$ vertex model are conveniently encoded in the
 $\check{R}$
matrix. This operator acts in a space $\left(C^{k+1}\right)^{\otimes 2}$, and
its elements are the Boltzmann weights of the corresponding configurations
(figure 1). An important technical point is that several choices of weights are
possible that are obtained by {\bf gauge} transformations. Such transformations
keep the Yang Baxter equation satisfied. Moreover they do not affect the
torus partition function of the system. There is a  special transformation that
 ensures commutation
of the $\check{R}$ matrix with the generators of $U_{q}sl(2)$, and in the
follow
   ing we
always consider this $\check{R}$ matrix as giving the weights of the model.
When
    the
$\check{R}$ matrix is in the commutant of $U_{q}sl(2)$, it satisfies some nice
a
   lgebraic
properties that define for instance in the $j=1/2$ case the {\bf Temperley
Lieb} algebra. An advantage of focusing on these algebraic properties is that
other representations can then be used, which  make the physics of the
problem more transparent.

 Consider for instance the $j=1/2$ case. The corresponding $6$
vertex model has a $\check{R}$ matrix (called r in the following) that writes
\begin{equation}
r=I+f(u)e
\end{equation}
where
\begin{equation}
q=\mbox{exp}(i\gamma),\gamma\in[o,\pi]
\end{equation}
$u$ is the spectral parameter, and
\begin{equation}
f(u)=\frac{\mbox{sin}u}{\mbox{sin}(\gamma-u)}
\end{equation}
The operator $e$ writes in the usual spin $j=1/2$ basis
\begin{equation}
e=\left(\begin{array}{cccc}
0&0&0&0\\
0&q^{-1}&-1&0\\
0&-1&q&0\\
0&0&0&0
\end{array}\right)\label{eq:e}
\end{equation}
and satisfies the equations
\begin{eqnarray}
e_{i}^{2}&=&\left(q+q^{-1}\right)e_{i}\nonumber\\
e_{i}e_{i\pm 1}e_{i}&=&e_{i}\nonumber\\
\left[e_{i},e_{j}\right]&=&0,|i-j|>1
\end{eqnarray}
that define the Temperley Lieb algebra. In these equations, $e_{i}$ is meant
for the operator that acts as (\ref{eq:e}) in the $i^{th}$ and $i+1^{th}$
copies
    of
$C^{2}$, and identity otherwise.

 It is however convenient to think of another
representation provided independently by knot theory ideas \cite{K87},
 and by polymer and
percolation problems. We notice that the Temperley Lieb
algebra can be represented
graphically as in figure 2. The operators act now in a space of lines, the
identity just  propagates the lines straight, while insertion of the
Temperley Lieb matrix $e$ attaches two neighbouring incoming and outgoing
 lines. Every time a closed loop is formed it
is assigned the value \footnote{In the following, $[x]_{q}$ denotes as usual
the q-analog, equal to $(q^{x}-q^{-x})/(q-q^{-1})$}
$[2]_{q}=q+q^{-1}$. We can therefore think of the $6$ vertex
model as a {\bf loop} model. There are now only two local configurations,
represented in figure 3, which we call (this notation will be explained later),
$I,x$, with respective weights $1,f(u)$. A very important remark is that on a
torus, the
partition function of the loop model equals that of  the original $6$ vertex
model {\bf if and only if the non contractible loops get a weight $2$ instead
of $[2]_{q}$}.

 The price to pay for having now only two local vertices is
that the computation of the weight of a configuration requires in general a
non local information, namely, when do pieces of lines attach together to form
a
loop, that should then be assigned a weight $[2]_{q}$. We recall
 that the loop evaluation used here is the geometrical
equivalent of the Markov trace \cite{W89}. There is however
 a magic point
where this non local information is not needed, it is obviously when
$[2]_{q}=1$. This corresponds to $q=exp(i\pi/3)$,  $g=1-1/3=2/3$, ie it is the
appropriate value for the $N=2$ point on the $6$ vertex (Gaussian) line. Notice
that at that point, the free energy can be evaluated in a trivial way: every
vertex has two possible states, so in the large volume limit where we can
neglect the special weight of non contractible loops we have
\begin{equation}
f=\mbox{log }2
\end{equation}
This point was actually studied in \cite{S91} from the point of view of twisted
$N=2$, and connected with the percolation problem. However,
 it does not seem that
one can consider the other values of $k$ in the series as representing some
multicritical percolation points. The situation will be different for polymers,
see below.

\subsection{Symmetrizers and Fusion}

The $\Gamma_{k}$ vertex model can be obtained from the $6$ vertex model by
projecting $k$ spins $j=1/2$ onto a spin $k$ representation of $U_{q}sl(2)$
\cite{KRS81,DJMO86}.
 The necessary
object for doing that is the $q$ symmetrizer, which has been widely studied
\cite{DWA88,We88,SA91}. It
can be expressed using  Temperley Lieb operators as follows. First recall that
\begin{eqnarray}
P_{0}&=&\frac{1}{q+q^{-1}}e\nonumber\\
P_{1}&=&{\cal S}_{2}=1-P_{0}
\end{eqnarray}
and introduce
\begin{equation}
s=q^{-1}P_{1}-qP_{0}
\end{equation}
Then one has
\begin{equation}
{\cal
S}_{k}=\frac{q^{k(k-1)/2}}{[k]_{q}!}\sum_{\sigma}q^{-|I_{\sigma}|}\prod_{i\in
I_{\sigma}}s(i)
\end{equation}
where the sum runs over permutations of $k$ objects. Decomposing $\sigma$ as a
product of transpositions $\tau_{i,i+1}$ gives the set of indices $I_{\sigma}$,
of minimal total number $|I_{\sigma}|$. The symmetrizers satisfy some simple
induction relations that make their determination easy. Now notice that ${\cal
S}_{2}$ can be represented graphically \cite{K90} as in figure 4. The property
$P_{0}P_{1}=0$ translates into the fact that strands always have to pass
through ${\cal S}_{2}$, and cannot make a U turn. This property generalizes to
any $k$.

Now the fusion procedure for getting the Boltzmann weights of say the
$\Gamma_{2}=44$ vertex model can be implemented in a variety of ways. We follow
here the analysis of \cite{DWA88,SA91}. One finds in these references the proof
that
\begin{equation}
\check{R}_{j=1}={\cal S}_{2}{\cal
S}_{2}r_{2}(u-\gamma)r_{3}(u)r_{1}(u)r_{2}(u+\gamma)
{\cal S}_{2}{\cal S}_{2}\label{eq:r2}
\end{equation}
This a natural generalization of the well known formula used in conformal field
theories or knot theory, that holds in the $u\rightarrow i\infty$ limit. It
corresponds to the picture of figure 5. Formula (\ref{eq:r2})
generalizes easily to give the
$\check{R}$ matrix of higher $k$ vertex models as well. One finds
\begin{eqnarray}
\check{R}_{j=k/2}&=&{\cal S}_{k}{\cal S}_{k}\left\{r_{k}[u-(k-1)\gamma]\ldots
 r_{2k-1}[u]\right\}\times
\left\{r_{k-1}[u-(k-2)\gamma]\ldots r_{2k-2}[u+\gamma]\right\}\nonumber\\
&\times&\ldots\times
\left\{r_{1}[u]\ldots r_{k}[u+(k-1)\gamma]\right\}
{\cal S}_{k}{\cal S}_{k}
\end{eqnarray}
We  remind the reader that $\check{R}$ has also an expression in terms of
the quantum projectors \cite{J86}
\begin{equation}
\check{R}(u)=P_{j=k}+\frac{y^{2}-q^{2k}}{1-y^{2}q^{2k}}P_{j=k-1}+\ldots
+\frac{y^{2}-q^{2k}}{1-y^{2}q^{2k}}\frac{y^{2}-q^{2(k-1)}}{1-y^{2}q^{2(k-1)}}
\ldots\frac{y^{2}-q^{2}}{1-y^{2}q^{2}}P_{j=0}
\end{equation}
where $y=exp(-iu)$.

\subsection{Geometrical Representation of the $\Gamma_{k}$ Vertex Models}

Let us start by discussing again the $j=1$ $\Gamma_{2}=44$ vertex model.
 We consider
the strand representation of the Temperley Lieb algebra introduced above, and
follow it through fusion. First it is clear that the $\check{R}$ matrix acts
now on {\bf bound states} ie on states that have two lines incoming each leg of
the four vertex. The symmetrizer acts on these lines, which is indicated
 by a transverse bar. Now we can compute the product (\ref{eq:r2}) by
picking one of the two terms in each of the $r$ factors, and drawing
accordingly either lines going straight or making some U turn. There are many
allowed such configurations, that are represented in figure 6\footnote{This
computation appears originally in some unpublished notes of Lou Kauffman
devoted to "Multi strands calculus" and cabling in knot theory. The present
formulation with insertion of symmetrizers leads to graphical rules for the
$SU(2)$ generalizations of the Jones polynomial (L.Kauffman, H.Saleur, work in
progress)}. However after
acting with the symmetrizers, only $3$ configurations remain, which we call
$1,x,x^{2}$, see figure 7. Their respective weights are
$W_{1},W_{x},W_{x^{2}}$, given by
\begin{eqnarray}
W_{1}&=&1\nonumber\\
W_{2}&=&f(u-\gamma)f(u+\gamma)f^{2}(u)\nonumber\\
W_{3}&=&f(u-\gamma)+f(u+\gamma)+2f(u-\gamma)f(u+\gamma)[f(u)+cos\gamma]
\end{eqnarray}
The isotropic point is, as for any $k$, obtained for $u=\gamma/2$, and one has
then
\begin{eqnarray}
W_{1}&=&1\nonumber\\
W_{2}&=&\frac{[2]_{q}}{1+[2]_{q}}\nonumber\\
W_{3}&=&1
\end{eqnarray}
so that the $W$ are invariant by a $90^{o}$ rotation. The three allowed
configurations can be written in the Temperley Lieb algebra as
\begin{eqnarray}
1&\leftrightarrow& {\cal S}_{2}{\cal S}_{2}\nonumber\\
x&\leftrightarrow& {\cal S}_{2}{\cal S}_{2}e_{2}{\cal S}_{2}{\cal
S}_{2}\nonumber\\
x^{2}&\leftrightarrow& {\cal S}_{2}{\cal S}_{2}e_{2}e_{1}e_{3}e_{2}
{\cal S}_{2}{\cal S}_{2}
\end{eqnarray}
As for the $6$ vertex model, although the interactions are much simpler in the
geometric picture (44 vertices replaced by 3), the computation of the total
Boltzmann weight of a configuration involves complicated evaluation of loops
weights. In our case, a single contractible
loop has weight $[2]_{q}$, while a
symmetrized contractible double loop (see figure 8) has weight $[3]_{q}$. A
single non contractible loop still has weight 2. The value of $q$
at which one can forget the loop weights, and the geometrical model becomes
local\footnote{The model is truly local for an infinite system only. In a
finite geometry, care has to be taken with non contractible loops. They however
, for the present regime, are negligible for the evaluation of thermodynamic
 quantitites}
 is given by
\begin{equation}
[3]_{q}=1,\ q=exp(i\pi/4)
\end{equation}
There are various ways of proving this assertion. One is to argue that since
the q dimension of the spin one representation, which is the fundamental
representation to consider in this case, is equal to one, the full Hilbert
space onto which the transfer matrix is acting has also a q dimension one,
irrespective of the number of spin one representations that are considered.
Therefore all quantum symmetric observables act trivially onto this space, and
there cannot be any non local contribution. This can be made more precise, but
is not very geometrical. Another proof is elementary and uses the very
geometrical definition of the model. Consider a typical "fat" graph appearing
in
the computation of the partition function of a large system (see figure 9).
Isolate in this graph an external "ear", and expand the symmetrizers that sit
on the two branches. One gets the pictorial equation
represented in figure 10. The first term has prefactor
\[
[2]_{q}-\frac{2}{[2]_{q}}=\frac{[3]_{q}-1}{[2]_{q}}
\]
that vanishes precisely at the point $q=\mbox{exp}(i\pi/4)=\mbox{exp}
(i\pi/(k+2))$.
 At this point we can therefore simply cut each ear to evaluate the value of
the graph, counting a factor $1/[2]_{q}^{2}$ for each suppressed ear.
By repeating inductively this procedure we end with a graph that has no ear and
therefore looks as the one of figure 11. We can decimate the loops using the
pictorial equation of figure 12 to get a factor $[2]_{q}-1/[2]_{q}$ per loop,
, which equals $1/[2]_{q}^{2}$ in that present case. The last loop gets a
factor $[3]_{q}$. Therefore the total weight of the
graph turns out to be
\begin{equation}
w_{\mbox{graph}}=\left(\frac{1}{[2]_{q}}\right)^{\mbox{area}-1}
\end{equation}
Now notice that
\begin{equation}
\mbox{area}=\mbox{number of vertices of type } x +1
\end{equation}
Therefore we can evaluate the partition function by forgetting about the
weights of contractible loops and multiplying the weights of each vertex by a
factor
\begin{equation}
\lambda_{1}=1,\lambda_{2}=1/[2]_{q},\lambda_{3}=1
\end{equation}
It is also interesting to consider the correspondence between the geometrical
vertices and the expansion of the $\check{R}$ matrix in terms of projectors.
One finds
\begin{eqnarray}
1&\leftrightarrow& P_{j=0}+P_{j=1}+P_{j=2}\nonumber\\
x&\leftrightarrow&
\left([2]_{q}-\frac{2}{[2]_{q}}\right)P_{j=1}+\left([2]_{q}-
\frac{1}{[2]_{q}}\right)P_{j=0}\nonumber\\
x^{2}&\leftrightarrow& \left([2]_{q}^{2}-1\right) P_{j=0}
\end{eqnarray}

The extension of these results to higher values of $k$ is easy. In general, the
legs of the vertices carry bound states made of $k$ symmetrized lines. There
remains only $k+1$ vertices, that can be labelled $1,x,\ldots,x^{k}$ see figure
13. These vertices correspond to interactions that are once again more
conveniently written graphically. We draw as in  figure 14 a square with each
face carrying an $e$ matrix with the appropriate label, and multiply the $e$'s
from top to bottom, and inside a given line, from left to right.
The precise form of the weights is not needed here (it involves use of quantum
$6j$ symbols\cite{KS91}). They have the
property of being symmetric under $90^{o}$ rotation for $u=\gamma/2$. At his
point one has therefore the symmetry
\begin{equation}
W_{x^{i}}=W_{x^{k-i}}
\end{equation}
The symmetrized contractible $k$ loops acquire a weight $[k+1]_{q}$ that
becomes equal to one for
\begin{equation}
q=exp(i\pi/k+2)\label{eq:spq}
\end{equation}
At this point one can show that the geometrical model becomes local, with
weights renormalized by the factors\cite{KS91}
\begin{equation}
\lambda_{x^{i}}=\frac{1}{[i+1]_{q}}
\end{equation}
Notice that at the special point (\ref{eq:spq}) one has
\begin{equation}
\lambda_{x^{i}}=\lambda_{x^{k-i}}
\end{equation}
Notice that, as in the $k=1$ case, the free energy can be simply evaluated at
the $N=2$ point of the $\Gamma_{k}$ vertex model. Since each vertex can be in
one of the $k+1$ allowed states independently of its neighbours, one gets in
the large volume limit where the effect of non contractible loops is negligible
\begin{equation}
f=\mbox{log}\left(\sum_{i=1}^{k+1}\lambda_{x^{i}}W_{x^{i}}\right)
\end{equation}

\subsection{Torus partition functions}

The loop model introduced above being equivalent to the $\Gamma_{k}$ vertex
model without charges at infinity, it is easy to derive its continuum limit
using the generalized Coulomg gas mapping. Following the results of
\cite{DFSZ88} one has
\begin{equation}
g=\frac{1}{k}-\gamma=\frac{2}{k(k+2)}
\end{equation}
The coupling between the parafermions and free bosonic field quantum numbers is
built in in the model. The only remaining point concerns the allowed range of
$m,m'$. Because we want the $x^{i}$ to be true operators acting in a transfer
matrix formalism, we need the square lattice at hand to be described diagonally
as in figure 15. There is therefore an even number of edges crossed when
describing one of the generators $\omega_{1},\omega_{2}$, so that following
\cite{DFSZ88}, $m,m'$ are integers. Formula (\ref{eq:z})  therefore follows.

\subsection{Lattice Candidates For Chiral Primaries}

We now present some arguments that the $x^{i}$ operators, that become local
operators at the magic point $q=exp(i\pi/k+2)$ for the $j=k/2$,
$k$ strands case, are
lattice candidates for the chiral primaries \cite{LVW89}. Of course as lattice
operators, the $x^{i}$ have a left and a right moving part, that behave
 identically. It is more proper to identify their lattice algebra with the
direct product of left and right chiral rings of the continuum limit. For
simplicity we think of the left moving sector only in the following.

It is straightforward to perform the multiplication $x^{i}.x^{j}$. For
this one simply connects all the outgoing lines of a $x^{j}$ vertex to the
incoming lines of an $x^{i}$ one, and expands the intermediate symmetrizers
(figure 16). The
result takes the form
\begin{equation}
x^{i}.x^{j}=\sum_{l=0}^{min(i+j,k)}c_{l}x^{l}\label{eq:prod}
\end{equation}
The precise form of the $c_{l}$ coefficients is not needed in the following.
The important point is that the right member {\bf truncates} and does not
contain terms with exponents greater than $i+j$. The reason for this is not
totally obvious. To form a term $x^{l}$ requires the introduction of $l$
matrices
$e_{k}$  (figure 17). On the other hand, $x^{i}$ and $x^{j}$ contain
respectively $i$ and $j$ such terms, while the symmetrizers do not contain any.
Therefore there are not enough $e_{k}$ terms to form an $x^{l}$ when
 $l>i+j$ in the
product. The fact that $i+j$ cannot exceed $k$ is simply because we start with
$k$ strands on each leg, so we cannot connect more than $2k$ of them.

Now  we can try to transform such a product into a meaningfull expression
in the continuum limit. For convenience we suppose that
we work on a cylinder with
some strands states propagating in the time direction (figure 18).
 After the conformal mapping from the plane to
the torus, fields become operators acting on the strands. Due to the conformal
factors, these operators have matrix elements that scale like
$L^{-h-\overline{h}}$, where $L$ is the radius of the cylinder.
 On the other hand the discrete operators $x^{i}$
introduced above have matrix elements that scale like $1$. Therefore we define
rescaled operators
\begin{equation}
X_{L}^{i}=L^{-2h_{i}}x^{i}\label{eq:regprod}
\end{equation}
so the product (\ref{eq:prod}) reads now
\begin{equation}
X_{L}^{i}.X_{L}^{j}=\sum_{l=0}^{i+j}c_{l}L^{2(h_{i+j-l}-h_{i}-h_{j})}X_{L}^{l}
\end{equation}
This equation has a (non trivial) limit when $L$ becomes large if and only if
\begin{equation}
h_{i+j}=h_{i}+h_{j}\label{eq:weight}
\end{equation}
In that case (\ref{eq:regprod}) becomes in the large $L$ limit
\begin{equation}
X^{i}.X^{j}=X^{i+j}\mbox{ for }i+j\leq k,\ 0\mbox{ otherwise}
\end{equation}
This algebra  is precisely the chiral ring \cite{VW89,M89},
 while the condition (\ref{eq:weight})
occurs as well in the study of continuum $N=2$ theories when one requires the
short distance expansion to reproduce the chiral ring. One finds then
\begin{equation}
h_{x^{i}}=i/2(k+2)
\end{equation}
Another point concerns the  $U(1)$ charge of the $x^{i}$
operators. Following \cite{S91} let us define it as proportional to the maximum
number of non contractible loops (in the time direction) the insertion of
$x^{i}$ can create. One finds
\begin{equation}
Q_{x^{i}}\propto i
\end{equation}
again a result known to hold for $N=2$ theories where
\begin{equation}
Q_{x^{i}}=i/(k+2)
\end{equation}
It is as well possible to consider the $x^{i}$ as antichiral primaries. They
satisfy the same algebra, but we define their  charge as
 proportional to minus the maximum number
of non contractible loops (in the time direction) their insertion can make
contractible (figure 19).
Notice also that the $90^{o}$ rotation
formally correspond to the so called Poincare duality \cite{VW89,M89}.
Finally it is clear that $I$
behaves indeed exactly as the identity in the transfer matrix formalism, and
therefore has vanishing conformal weight. We discuss the conformal weights of
the other $x^{i}$ fields in the next section.

\subsection{Discussion}

First we discuss the case $j=1/2$. We therefore have contractible contours with
weight $[2]_{q}=1$ for $q=\mbox{exp}i\pi/3$, non contractible contours with
weight two, and two vertices of interaction,
$I,x$ with the same weight one. We consider the model on a cylinder.
 It is easy to isolate the doubly periodic sector by considering for a
while configurations where the non contractible loops also have a weight one.
In that case the partition function is obviously equal to a constant,
 in agreement
with $Z_{PP}=Z_{\tilde{R}}=2$. This last value is merely an overall
normalization problem when taking the continuum limit in the Coulomb gas
mapping. It can be found by noticing that the loop configurations so isolated
just compute the partition function of the $A_{2}$ model, which corresponds in
turn to taking yet another representation of the Temperley Lieb algebra of
solid on solid type, with elements for the face configuration of figure 20
\begin{equation}
e_{l_{1}l_{2}l_{3},l_{1}l'_{2}l_{3}}=\delta_{l_{1}l_{3}}
\frac{\sqrt{[l_{2}]_{q}[l'_{2}]_{q}}}
{[l_{1}]_{q}}
\end{equation}
We can therefore picture the Ramond ground state as a
frustration line running accross the cylinder in time direction, such that
every loop crossing it gets a weight $1$ instead of $2$. In the Coulomb gas
mapping, this corresponds to having a pair of electric charges at the
extermities of the cylinder, with value $e_{0}=\pm1/3$. This has dimension
$h=e_{0}^{2}/4g=1/24$ as expected. Let us now consider the full theory, with
non contractible loops having weight 2\footnote{Unfortunately we do not know
what is the microscopic definition of all the sectors in that case. In that
respect, the polymer situation described in the next section is more favorable}
. Because insertion of $I$ acts
identically on the propagating strands, its dimension is certainly
\begin{equation}
h_{I}=0
\end{equation}
Consider now the operator $x$. It is reasonable to assume that the connected
correlation function with its antichiral conjugate is obtained by creating a
pair of non contractible loops where $x$ is inserted, and destroying it when
the conjugate is inserted: in other words by selecting
 the configurations where a loop
connects the two insertions as in figure 21. It is well known how to evaluate
this dimension in the Coulomb gas mapping \cite{N84}. The corresponding
operator
    is
magnetic with charge $m=1$, and therefore
\begin{equation}
h_{x}=\frac{g(m=1)^{2}}{4}=\frac{1}{6}
\end{equation}
as expected. This dimension is also what would be observed if we were to
compute the correlation function in the antiperiodic (NS) sector.

This picture essentially generalizes to the higher $k$ case. The doubly
periodic sector is reproduced by using the restricted solid on solid model
representation of the Temperley Lieb algebra for $A_{k+1}$. The Witten index is
$k+1$ corresponding to the number of nodes of the diagram. The Ramond ground
state is obtained by giving to the non contractible loops accross the cylinder
a weight $[2]_{q}$. Finally if we admit the likely hypothesis that connected
correlation functions of $x^{i}$ operators and their conjugates
 are described by $i$ loops joigning
the two insertions we find, using the generalized Coulomb gas mapping, the
weights $h_{x^{i}}=i/2(k+2)$.

\section{D type invariants}

It is well known how the $D_{2+k/2}$ theories can be obtained from
the $A_{k+1}$ theories by a $Z_{2}$ orbifold. In our case it is easy to form
symmetric combinations of the diagrams $x^{i}$ and $x^{k-i}$. However we do not
know what the equivalent of the $y$ field \cite{VW89,M89} should be. Hints in
the identification of $D$ lattice models can be obtained from consideration of
the partition functions. Using once again the results of \cite{DFSZ88} one
finds first that the expression (\ref{eq:z}) generalizes to $D$ or $E$
invariants, but with the appropriate parafermions partition functions
(\ref{eq:zcoul}). Then since $D$ parafermions are themshelves orbifold of the
$A$ ones one has
\begin{eqnarray}
Z^{D_{2+k/2}}_{k}(r,s)&=&\frac{1}{2}\left[Z_{k}(r,s)+(-)^{r}Z_{k}(r,s+k/2)+
(-)^{s}Z_{k}(r+k/2,s)\right.\nonumber\\
&+&\left.(-)^{r+s}Z_{k}(r+k/2,s+k/2)\right]
\end{eqnarray}
The peculiar combination of sectors of the $\Gamma_{k}$ vertex model
reproducing the $D_{2+k/2}$ invariant can obviously be read from the above
equation. But it still lacks a natural interpretation.

\section{$K^{th}$ Critical Polymer Models and Twisted $N=2$ with Level $k$}

There is another, and physically more interesting, way of interpreting the
$N=2$ partition functions. Recall that the $\Gamma_{k}$ vertex models are
expected to have in fact {\bf two} phases. In the first one, the model
 is integrable
with Boltzmann weights given by the usual formulas \cite{KRS81,DJMO86},
 and coupling
constant as above
\begin{equation}
g=\frac{1}{k}-\gamma,\ q=e^{i\gamma},\ \gamma\in\left[0,1/k\right]
\end{equation}
The second phase is not very well known except for $k=1$. It is supposed to be
obtained by adding some vacancies. There are however some strong arguments
\cite{DFSZ88} that the
corresponding renormalized coupling constant is
\begin{equation}
g=\frac{1}{k}+\gamma
\end{equation}
In the case $k=1$ the first regime corresponds to the critical Potts model with
$Q=\left(q+q^{-1}\right)^{2}$, the second regime to the critical  $O(n)$
model with  $n=q+q^{-1}$ . For higher $k$ it is likely that the second regime
de
   scribes a
sort of multicritical $O(n)$ model with in particular the formation of {\bf
boun
   d
states}, ie up to $k$ lines colliding on a same edge, allowed.
Now choose $\gamma=1/2$. This value corresponds to $n=0$ ie to $k^{th}$
critical {\bf polymers}.
The associated coupling constant is
\begin{equation}
g=\frac{k+2}{2k}
\end{equation}
dual to the coupling constant for the
vertex model that we discussed
in the preceding section. In the case of polymers, we can however generalize
the discussion of \cite{S91} for the dense and critical case, and
provide a natural geometrical interpretation of the various quantities so far
encountered. The natural point of view here is to consider {\bf twisted} $N=2$
theories \cite{EY90}.
 Recall that twisting the theory is obtained by adding a term
$\frac{1}{2}\partial J$ to the stress energy tensor. The new central charge
vanishes. The partition function of the twisted model is however the same as
the one of the untwisted model, so twisting merely amounts to a change of point
of view. The former ground state of the Ramond sector is now considered as the
true $SL_{2}$ invariant ground state, while the former true
 ground state is now considered as a
state with negative dimension. Notice also that, because of twisting, the
fermio
   ns
acquire integer dimensions and therefore have identical boundary conditions for
the cylinder and the plane.

\subsection{Polymers Partition Functions}

First we notice that for polymers also the bulk free energy is easy to
evaluate. Polymers occupying a vanishing fraction of the available space one
has $f=0$, as is usual for $n\rightarrow 0$ limit models.

As in \cite{S91} we think of the $n\rightarrow 0$ limit as obtained by
associating to each polymer loop a bosonic or a fermionic variable. In this way
all contractible loops on a torus disappear, while non contractible loops can
get a weight zero or two depending on their winding and the fermions boundary
conditions (figure 22). One has, if the labels A or P refer to these discrete
fe
   rmions
boundary conditions

{\bf Definition}

${\cal Z}_{AP(\mbox{resp.}PA,\mbox{resp.}AA)}=$ Sum over configurations with an
{\bf even} number of non contractible $k^{th}$ critical polymers of total
length ${\cal L}$, that cross
$\omega_{2}$ (resp.$\omega_{1}$,resp.$\omega_{1}+\omega_{2}$) an {\bf odd}
number of times, with weight $2^{\mbox{number of polymers}}\mu^{-{\cal L}}$
$\times$ the Boltzmann weights induced by the $k^{th}$ critical interactions

\smallskip

where $\mu$ is the appropriate inverse connectivity constant. One can also
define the manifestly modular invariant quantity
\[
{\cal Z}^{e}=\left\{\mbox{Sum over configurations with an {\bf even} number of
 non contractible }k^{th}\mbox{ critical}\right.
\]
\[
\mbox{polymers of total  length }{\cal L}\mbox{ with
weight }2^{\mbox{number of polymers}}
\]
\begin{equation}
\left.\mu^{-{\cal L}}\times\mbox{the Boltzmann
weights induced by the }k^{th}\mbox{critical interactions}\right\}
\end{equation}
One checks easily from these definitions that\footnote{Recall that the two
winding numbers of a non contractible loop are coprimes}
\begin{equation}
{\cal Z}^{e}=\frac{1}{2}\left({\cal Z}_{AP}+{\cal Z}_{PA}+{\cal Z}_{AA}-{\cal
Z}_{PP}\right)
\end{equation}
Notice that projection on odd fermion number occurs naturally
on geometrical grounds.

  On the other hand the analysis with the generalized lattice
 Coulomb gas is such that topological properties (magnetic defects,
electric charges, winding numbers...) do not depend on the peculiar $k^{th}$
critical regime. We can therefore use the same expressions as the ones in
\cite{S91} to express in the continuum limit
\[
Z_{PP}=\sum_{r,s=0,\ldots,k-1}Z_{k}(r,s)\sum_{\begin{array}{c}
m=r\mbox{ mod} k\\
m'=s\mbox{ mod }k
\end{array}}\left(-\right)^{m\wedge m'}Z_{mm'}
\]
\[
Z_{AP}=\sum_{r,s=0,\ldots,k-1}Z_{k}(r,s)\sum_{\begin{array}{c}
m=r\mbox{ mod} k\\
m'=s\mbox{ mod }k
\end{array}}\left(-\right)^{m}\left(-\right)^{m\wedge m'}Z_{mm'}
\]
\[
Z_{PA}=\sum_{r,s=0,\ldots,k-1}Z_{k}(r,s)\sum_{\begin{array}{c}
m=r\mbox{ mod} k\\
m'=s\mbox{ mod }k
\end{array}}\left(-\right)^{m'}\left(-\right)^{m\wedge m'}Z_{mm'}
\]
\begin{equation}
Z_{AA}=\sum_{r,s=0,\ldots,k-1}Z_{k}(r,s)\sum_{\begin{array}{c}
m=r\mbox{ mod} k\\
m'=s\mbox{ mod }k
\end{array}}\left(-\right)^{m+m'}\left(-\right)^{m\wedge m'}Z_{mm'}
\end{equation}
One checks that each of these expressions coincides respectively with the known
partition functions of the $N=2$ theories. In that case we therefore have found
a geometrical interpretation for all the sectors of the theory.
Let us emphasize again that due to twisting, the
fermions acquire integer dimensions and therefore have the same boundary
conditions in the plane and on the torus\footnote{In the preceding paper
\cite{S91} we referred to R and NS as the boundary conditions in the plane}.
 Summing
these four contributions one gets
\begin{equation}
Z^{e}=Z_{gc}\left(\frac{k+2}{2k}\right)\mbox{, for $k^{th}$ critical polymers}
\end{equation}
As in the $k=1$ case, this partition function is not consistent physically
because it does not contain the identity (for the twisted theory), as expected
since we projected on odd fermion number. The physical polymer partition
function is rather obtained by projection on even fermion number
\begin{equation}
{\cal Z}^{phy}={\cal Z}^{e}+k+1\rightarrow Z^{e}+k+1
\end{equation}
The ground states in the twisted theory describe simply configurations with no
polymers. That they appear with some multiplicity occurs probably because of
the additional degrees of freedom introduced to make the polymers multicritical
(vacancies and their generalizations \cite{Bernard82}).
As in the $k=1$ case we notice that R and NS correctly describe
observables
with an {\bf even} number of polymers only. The full polymer theory would need
i
   n
addition the consideration of a sector with $Z_{4}$ twists \cite{S91}
, which is easily
studied via the spectral flow. Also we recall that the BRS cohomology used to
obtain topological theories out of twisted $N=2$ \cite{EY90}
is {\bf not} the right procedure to
extract physical states from the polymer point of view. On the contrary,
$Q_{BRS}$ turns out to be the operator that creates polymers out of the vacuum.

\subsection{Some Polymer Exponents and the Flory Formula}

The most important quantity is the exponent $\nu$ that controls the
mean size of polymers. Considering a polymer loop of ${\cal L}$ monomers one
has asymptotically
\begin{equation}
\left<R_{G}^{2}\right>\propto{\cal L}^{2\nu}
\end{equation}
where $R_{G}$ is the radius of gyration. If we introduce the 2 legs polymer
operator, of conformal weights $h=\overline{h}$,
 whose correlations are defined by summing over configurations of a
polymer loop attached at two different points, one has, by standard scaling
arguments
\begin{equation}
\frac{1}{\nu}=2-2h\label{eq:nu}
\end{equation}
This operator was identified in \cite{S91} to be the first non trivial operator
in the periodic sector ($X$ in the Landau Ginzburg picture),
 with weight after twisting given by
\begin{equation}
h=\frac{1}{k+2}
\end{equation}
Using formula (\ref{eq:nu}) one finds
\begin{equation}
\nu=\frac{k+2}{2(k+1)}\label{eq:exp}
\end{equation}
These exponents have already appeared in the literature and they are known
as the Flory exponents \cite{F71}
for $k^{th}$ multicritical polymers (in our conventions, ordinary polymers are
$k=1$). The Flory formula is obtained by making crude, and notably wrong
\cite{D76},
assumptions for the free energy of a polymer. One usually writes $F$ as the sum
of an elastic and energetic contributions, and minimizes with respect to say
${\cal L}$ at $R_{G}$ fixed to obtain $\nu$. The elastic contribution is argued
to be $F_{el}\propto R_{G}^{2}/{\cal L}$, and the energy contribution for a $k$
multicritical point $F_{en}\propto R_{G}^{d} ({\cal L}/R_{G}^{d})^{k+1}$ where
$d$ is the space dimension \cite{DG85}. The argument for this last expression
mimics field theory where for ordinary polymers, the
$\phi^{4}$ term dominates, corresponding to two pieces of polymers that come in
contact (figure 23), for tricritical polymers the $\phi^{6}$ term dominates
corresponding to three pieces of polymers coming in contact\cite{Ber82}...
Writing
therefore
\begin{equation}
F\propto\frac{R_{G}^{2}}{{\cal L}}+R_{G}^{d}\left(\frac{{\cal
L}}{R_{G}^{d}}\right)^{k+1}
\end{equation}
one finds after minimization
\begin{equation}
\nu=\frac{k+2}{2+dk}\label{eq:flory}
\end{equation}
One could not really take this formula seriously based only on the Flory
"derivation". However in the case $k=1$ it is in fact exact for $d=1,2,4$ and
very close to the numerical estimates for $d=3$. $\nu=1/2$ for $d=4$ which one
can therefore correctly identify as the  upper critical
dimension. Formula (\ref{eq:flory}) should not be applied above the critical
dimension, where the exponent  $\nu$ is expected to remain equal to $1/2$.
 Some
explanations of approximate validity of Flory's formula have been proposed
\cite{B89}.
What is fascinating however is that it can be {\bf exact} in most cases.
 So far higher $k$'s
 had not been really considered. We notice however that if we
put $d=2$ in (\ref{eq:flory}) we recover the exponents (\ref{eq:exp}). This
shows they are also exact for the peculiar kind of multicritical polymers at
hand, and strongly suggests that the Flory formula can be exact because of {\bf
non renormalization theorems} due to the hidden $N=2$ supersymmetry in
polymers. As a final argument in that direction, notice that $\nu$ in
(\ref{eq:flory}) becomes equal to $1/2$ for
\begin{equation}
d=2+\frac{2}{k}
\end{equation}
which is precisely the upper critical dimension for $k^{th}$ critical $N=2$
Landau Ginzburg theories.

\subsection{Renormalization Group Flow}

        It is  worth considering the problem from the point of view of
renormalization group flow \cite{Z86,KMS89}. We remark that because
twisting is only a change of point of view on the system, the flow results
established for the untwisted case should hold also for polymers. Indeed
perturbing for instance by $X$ the $k$ theory means shifting the geometrical
weight
of the polymer to  $(\mu+\delta\mu)^{-{\cal L}}$, independently of whether we
consider the ground state as the state without polymers, or the state with
large non contractible loops that have weight two.

For the $N=2$ series, the  "minimal" supersymmetry preserving flow is
a flow from the  theory with $k$ to the next theory with $k-2$
 ($X^{k-1}$ is redundant). It is generated by the superpartner of
the top component
of the chiral primary field $X^{k}$, with expression in the generalized Coulomb
gas, where $\psi_{1}$ is the parafermionic field of lowest dimension,
\begin{equation}
\Phi_{\mbox{pert}}=\psi_{1}\overline{\psi}_{1}exp\left(\frac{2i}{k}\phi
(z,\overline{z})\right)
\end{equation}
and dimension $h=\overline{h}=\frac{k+1}{k+2}$.
 After twisting, since $G^{+}$ acquires
weight one, this field has the same dimension as $X^{k}$ in the twisted
theory, which can also be computed by adding the proper charge at infinity for
the field $\phi$ in the generalized Coulomb gas. One finds
\begin{equation}
h=\frac{k}{k+2}
\end{equation}
This perturbation is still integrable in the twisted theory as discussed in
\cite{EY90}. The above dimension provides the value of the exponent $\nu_{u}$
for the multicritical polymers at hand \cite{DS87bis}
\begin{equation}
\nu_{u}=\frac{k+2}{4}
\end{equation}
and therefore the crossover exponent
\begin{equation}
\varphi=\frac{\nu}{\nu_{u}}=\frac{2}{k+1}
\end{equation}
We recall that the renormalization group flow in the $N=2$ series is more
subtle than in the $N=0$ or  $N=1$ series. The analysis of the superpotential
$X^{k+2}+\lambda X^{k}$ can easily be done, at least naively,
 since there is only wave function
renormalization \cite{WR89}. However two points with different values of $k$
are
infinitely far away in the sense of \cite{CK90}, so the flow cannot be studied
in the conformal perturbation framework\footnote{From the polymer point of
 view, and owing to
experimental and numerical knowledge of multicritical polymer systems, there is
little doubt that such flow occurs}. This is because
the index takes value $k+1$ for the $k^{th}$ critical polymers, and therefore
has to jump by two units between the $k$ and the $k-2$ theory. As long as
$0<k<\infty$ one does not expect spontaneous breaking of supersymmetry.

The perturbation with the operator $X$ is also integrable \cite{FMV90},
 and physically
 interesting. Recall that $X$ was identified with the two polymer legs
operator, and therefore adding an $X$ perturbation is like changing the
geometrical weight of polymers. Two behaviours can
occur depending on the sign of the coupling. For the sign corresponding to
$\delta\mu>0$, the theory flows to a trivial fixed point. This describes
so small polymers that they disappear at large scale. For the sign
corresponding to $\delta\mu<0$, we expect physically the polymers to become so
large that they occupy a finite fraction of the available volume, with fractal
dimension exactly equal to two, that is an exponent $\nu=1/2$.
 This means the system should flow to the dense
polymers phase, that is described by an $\eta,\xi$ system \cite{S91}, and that
supersymmetry is {\bf spontaneously broken}. The
decoupling of the ground state has a very transparent physical interpretation.
 Indeed for  the state
without polymers ${\cal Z}\propto 1$, while as soon as one {\bf dense}
polymer is allowed
${\cal Z}\propto e^{f\times\mbox{area}}$ for f
 some non vanishing free energy. Notice that in the dense phase, the free
energy is now a non trivial number, which is not known exactly except in
special cases (hamiltonians walks on the Manhattan lattice for instance).

For $k=1$ (ordinary self avoiding polymers), the preturbations $X^{k}$ and $X$
coincide. The "bottom" of the series of multicritical points is therefore dense
polymers, where the supersymmetry is spontaneously broken.

\subsection{$k\rightarrow\infty$ limit}

        Another insight can be obtained
by considering the $k\rightarrow\infty$ limit. In that limit the exponent $\nu$
tends to $1/2$, the exponent of {\bf dense} polymers\footnote{$\nu=1/2$ is of
course also the exponent of brownian walks, but this is not apparently the same
problem, although some other features are common with the dense case.}. This
corresponds to polymers that are collapsed onto themshelves, due to strong
attractive interactions, or compression from the exterior. We therefore find
again dense polymers, which are
 indeed
expected to be the end point of  the multicritical polymers series when
 described in
the direct renormalization or $O(n),n\rightarrow 0$
 multicritical field theory. We
can here make the identification of the $k\rightarrow\infty$ limit
of our theories
with dense polymers very precise. Indeed it
was shown in \cite{S91} that dense polymers
are conveniently described by an $\eta,\xi$
system with central charge $c=-2$. This corresponds to breaking the symmetry
between bosons and fermions in say the free field representation of $N=2$
twisted theories. By summing over
various sectors, it was shown that the entire partition function of dense
polymers with an even number of non contractible loops on the torus is a
gaussian partition function
\begin{equation}
Z_{\mbox{dense}}=Z_{c}[1/2]
\end{equation}
where by $Z_{c}$ we mean an expression similar to $Z_{gc}$ but with the
partition functions of the $Z_{1}$ parafermions taken to be equal to one. On
the other hand it is possible to work out the $k\rightarrow\infty$ limit of the
partition functions (\ref{eq:z}).
 For this purpose notice first that the coupling
constant  goes to a finite value $g=1/2$ in that limit. For $k$ large
enough, the $m,m'$ excitations are rapidly damped out in the second summation
in the expression (\ref{eq:zcoul}) of the generalized
 Coulomb partition function, so we
can truncate this summation to $m=r,m'=s$. Moreover as $k$ becomes large, the
frustrations of the $Z_{k}$ model become negligible for the values of $r,s$
such that $Z_{rs}$ is not itself too small. One can therefore write
\begin{equation}
Z_{gc}\left(\frac{k+2}{2k}\right)\propto Z_{k}(0,0)Z_{c}[1/2],\mbox{ as
}k\rightarrow \infty
\end{equation}
Now such a result can also be established for the level $k$ Wess Zumino
partition functions which can be rewritten as a
generalized Coulomb partition function for a coupling $g=1/k$. On the other
hand, using reexpression of these same WZW partition functions in terms of 3
bosonic partition functions as worked out in \cite{JB} leads to
\begin{equation}
Z_{WZW}\propto\left(\frac{1}{\sqrt{Im\tau}\eta\overline{\eta}}\right)^{3}
\end{equation}
Combining these results we get finally
\begin{equation}
Z_{gc}\left(\frac{k+2}{2k}\right)\propto\left(\frac{1}{\sqrt{Im
\tau}\eta\overline{\eta}}\right)^{2} Z_{c}[1/2]\mbox{ as
}k\rightarrow\infty
\end{equation}
Therefore in the large $k$ limit we find indeed the dense polymer partition
function, up to a free bosonic field factor that helps maintaining central
charge at a value zero. As in the preceding study of the $X$
perturbation, one expects in the $k\rightarrow\infty$ that
spontaneous braking of supersymmetry will occur, leaving only the $\eta,\xi$
system with partition function $Z_{c}[1/2]$. Notice that the
untwisted model corresponding to the $k\rightarrow\infty$ limit is the
$c=3$ $N=2$ theory, which is well known to have vanishing index.
 Therefore dense polymers occur
both at the "top" and the "bottom" of the multicritical series.
That dense polymers appear both
in the $k\rightarrow\infty$ and in the low temperature phase
 is due to the existence
of two independent mechanisms for obtaining them. When $k\rightarrow\infty$,
more and more attractive interactions are added, so the polymer is forced to
collapse onto itself for internal reasons. On the other hand, the breaking
of symmetry of the $O(n),n\rightarrow 0$ model actually reduces the available
volume, so the polymer collapses for topological reasons \cite{DS87}.

\subsection{Ordinary theta point}

Can we identify one of the multicritical points discussed above
with
the usual theta or theta'\cite{DG85,DS87}\footnote{The distinction of theta and
theta' points is a technical matter that we cannot discuss here. See
\cite{DS87}
and subsequent comments and replies in Phys. Rev. Lett.}
point? We recall that the
theta point, which was solved in \cite{DS87} involves
 a set of attractions between
nearest and second nearest neighbours on the honeycomb lattice, with a
partition
function, for an even number of non contractible polymer chains, that was
supposed so far to equal
$Z_{c}[2/3]$. The point $k=6$ in our series has also $g=2/3$, but a rather
different operator content due to the $Z_{6}$ parafermions and the boundary
conditions couplings. The exponent $\nu=4/7$ is the same. For the theta point
it is expected, and rather well established numerically, that
the cross over exponent is
$\varphi=3/7$. This means that the perturbation physically identified as
changing the attraction between monomers has dimension
\begin{equation}
h=1-\frac{\varphi}{2\nu}=\frac{5}{8}
\end{equation}
This is the dimension of the field $X^{5}$ after twisting, while for $k=6$ the
theory has an $X^{8}$ potential. Under this perturbation the theory flows to
the $k=3$ theory however, not the $k=1$ theory! This can be interpreted in two
ways. We can of course believe that the present set of multicritical points has
nothing in common with the theta point, and that the possible concordance of
some exponents is an accident. We can also try to minimize the available set of
polymer systems by assuming that $k=6$ is indeed the theta point,
and that something was not understood in  \cite{DS87}. This would explain why
th
   e numerical analysis of the problem is
so difficult. Indeed in the present scheme, we can go from $k=6$ to $k=1$ with
a crossover exponent $\varphi=3/7$ at the $k=6$ point only by passing through
the $k=3$ point, with $\nu=5/8$. The mixture of these behaviours for the
lattice polymers may indeed lead to very confusing measurments. We summarize
some of the properties of the polymer multicritical series in figure 24.

\section{Conclusions}

In conclusion the two kinds of models we have introduced illustrate various
aspects of $N=2$ theories. The geometrical reformulation of the $\Gamma_{k}$
 vertex model leads to lattice analogs of the chiral ring.
The multicritical polymers are not precisely defined yet \footnote{We do not
know their exact Boltzmann weights, as is usually the case for multicritical
systems}. But they provide a nice interpretation of the role of bosonic and
fermionic degrees of freedom, a geometrical understanding of the boundary
conditions, and hopefully physical realizations of $N=2$ flow and spontaneous
supersymmetry breaking. For both models, the bulk free energy can be readily
obtained, without Bethe ansatz type computations.

This work opens many physical questions. Among them,
one would like to find an explicit $N=2$ formulation of
 the Flory approximations, which maybe could be generalized to higher
dimensions. Also, the series of multicritical  polymer points seems a good
place to apply the ideas of \cite{Zlast}. Notice that polymers can give rise to
experiments in two dimensions \cite{VR}, and therefore it should be possible to
observe in such systems consequences of $N=2$ supersymmetry.

On the more mathematical side, the study of interrelations
between the singularity structure and the geometrical lattice models should be
fruitful.

Finally, it seems from our study that the most natural description of
lattice models with
$N=2$ continuum limit is geometrical. This is maybe not totally unexpected. For
instance that the free energy can be obtained without
computation, which is the kind of nice  property one expects for a $N=2$ model,
seems to imply a geometrical setting  where the non trivial observables
 are defined by topological properties, for instance the connectivity in
polymers \cite{S91}. Maybe by pursuing this route one could indeed give lattice
interpretations of the magic numbers that have been computed for topological
theories \cite{V91,CV91}.

\bigskip

{\bf Acknowledgments}: We thank A.Leclair, P.Fendley, G.Moore, N.Read,
N.Seiberg
   ,
N.Warner and
A.Zamolodchikov for many useful discussions.

\pagebreak
{\bf Figure Captions}

\bigskip

Figure 1: The weights of vertices are encoded in the $\check{R}$ matrix.

Figure 2: Graphical representation of the Temperley Lieb algebra.

Figure 3: The two vertices in the geometrical reformulation of the 6 vertex
model.

Figure 4: Geometrical representation of the symmetrizer on two strands.

Figure 5: Geometrical representation of the fusion formula for $\check{R}$
matrices.

Figure 6: The various configurations that enter in the decomposition of the
$j=1$ fusion before applying the symmetrizers.

Figure 7: The three remaining configurations after application of the
symmetrizers.

Figure 8: The value of the double contractible loop is $[3]_{q}$.

Figure 9: A typical fat graph entering the computation of the partition
function.

Figure 10: Factorization rule for one "ear" of the diagram.

Figure 11: The same diagram after factorizing out the ears.

Figure 12: Factorization of "handles".

Figure 13: The four vertices of interaction for $k=3$.

Figure 14: Squares with  $e$ matrices on each face encode conveniently
 the algebraic form of the $x^{i}$ interaction terms. This figure corresponds
to the interaction of figure 17.

Figure 15: Diagonal propagation for the square lattice.

Figure 16: Graphical computation of $x^{2}$ for $k=2$.

Figure 17: Formulation of $x^{3}$ in the Temperley Lieb algebra.

Figure 18: Strands propagating on a cylinder

Figure 19: Insertion of $x$ reduces the number of non contractible loops by two

Figure 20: $e$ matrix in the solid on solid basis.

Figure 21: Configurations that contribute to the connected correlation
function for the $X$ operator.

Figure 22: If the wiggly line represents anti periodic boundary conditions for
fermions, a non contractible loop that crosses it once gets a factor $1+1=2$,
while a non contractible loop that crosses it twice gets $1-1=0$.

Figure 23: Multicritical polymers are obtained by adding multipieces
interactions.

Figure 24: Some features of the renormalization group flow in the multicritical
polymers series.

\pagebreak


\begin{thebibliography}{99}
\bibitem{KMS89}D.Kastor, E.Martinec, S.Shenker, {\sl Nucl. Phys.}, {\bf B316},
590 (1989)
\bibitem{VW89}C.Vafa, N.Warner, {\sl Phys. Lett.}, {\bf B218}, 51 (1989)
\bibitem{M89}E.Martinec, {\sl Phys. Lett.}, {\bf B217}, 431 (1989)
\bibitem{HW}P.S.Howe, P.C.West, {\sl Phys. LEtt.}, {\bf B223}, 377 (1989)
\bibitem{DFSZ88}P.Di Francesco, H.Saleur, J.B.Zuber, {\sl Nucl. Phys.}, {\bf
B300}, 393 (1988)
\bibitem{FS}D.Friedan, S.Shenker, in "Conformal Invariance and Applications to
Statistical Mechanics", C.Itzykson et al. Editors, World Scientific (1988)
\bibitem{GR87}G.von Gehlen, V.Rittenberg, {\sl J.Phys.}, {\bf A20}, 227 (1987)
\bibitem{YZ87}S.K.Yang, Y.B.Zheng, {\sl Nucl. Phys.}, {\bf B285}, 410 (1987)
\bibitem{Z86}Z.B.Zamolodchikov, {\sl Sov. Journal. Nucl. Phys.}, {\bf 44}, 529
(1986)
\bibitem{S91}H.Saleur,"Polymers and Percolation in Two Dimensions
 and Twisted N=2
supersymmetry", preprint YCTP-P38-91
\bibitem{W88}E.Witten, {\sl Comm. Math. Phys.}, {\bf 118}, 411 (1988);
E.Witten, {\sl Nucl. Phys.}, {\bf B340}, 281 (1990)
\bibitem{EY90}T.Eguchi, S.K.Yang, {\sl Mod. Phys. Lett.}, {\bf A5}, 1693 (1990)
\bibitem{PS80}G.Parisi, N.Sourlas, {\sl Jour. de Phys. Lett.}, {\bf 41}, L403
(1980)
\bibitem{F71}P.J.Flory, "Principles of Polymer Chemistry", Cornell University
Press, Ithaca (1971)
\bibitem{R88}F.Ravanini, {\sl Mod. Phys. Lett.}, {\bf A3}, 271, 397
(1988); D.Kastor, E.Martinec, Z.Qiu, {\sl Phys. Lett.}, {\bf B200}, 434 (1988);
J.Bagger, D.Nemeschansky, S.Yankielowicz, {\sl Phys. Rev. Lett.}, {\bf 60}, 389
(1988)
\bibitem{ZF85}A.B.Zamolodchikov, V.A.Fateev, {\sl Sov. J. Physics}, {\bf JETP
62}, 215 (1985)
\bibitem{GQ87}D.Gepner, Z.Qiu, {\sl Nucl. Phys.}, {\bf B285}, 423 (1987)
\bibitem{RY87}F.Ravanini,S.K.Yang, {\sl Phys. Lett.}, {\bf B195}, 202 (1987);
Z.Qiu, {\sl Phys. Lett.}, {\bf 198B}, 497 (1987); D.Gepner, {\sl Nucl. Phys.},
{\bf B296}, 757 (1987); V.K.Dobrev, A.C.Ganchev, {\sl Mod. Phys. Lett.}, {\bf
A3}, 127 (1988)
\bibitem{DFSZ}P.di Francesco, H.Saleur, J.B.Zuber unpublished
\bibitem{Y88}S.K.Yang, {\sl Phys. Lett.}, {\bf B209}, 242 (1988)
\bibitem{K87}L.Kauffman, "On Knots", Princeton University Press (1987); "Knots
And Physics", World Scientific (1991)
\bibitem{W89}M.Wadati, T.Deguchi, Y.Akutsu, {\sl Phys. Rep.}, {\bf 180} (1989)
and references therein
\bibitem{KRS81}P.Kulish, N.Y.Reshetikhin, E.K.Sklyanin, {\sl Lett. Math.
Phys.}, {\bf 5}, 393 (1981)
\bibitem{DJMO86}E.Date, M.Jimbo, T.Miwa, M.Okado, {\sl Lett. Math. Phys.}, {\bf
12}, 209 (1986)
\bibitem{DWA88}T.Deguchi, M.Wadati, Y.Akutsu, {\sl J.Phys. Soc. Jap.}, {\bf
57}, 1905 (1988)
\bibitem{We88}H.Wenzl, {\sl Invent. Math.}, {\bf 92}, 349 (1988)
\bibitem{SA91}H.Saleur, D.Altschuler, {\sl Nucl. Phys.}, {\bf B354}, 579
(1991)
\bibitem{K90}L.Kauffman, "Knots, Spin Networks and 3 Manifold Invariants",
preprint
\bibitem{J86}M.Jimbo, {\sl Comm. Math. Phys.}, {\bf 102}, 547 (1986)
\bibitem{KS91}W.M.Koo, H.Saleur, in preparation
\bibitem{LVW89}W.Lerche, C.Vafa, N.P.Warner, {\sl Nucl. Phys.}, {\bf B324}, 427
(1989)
\bibitem{N84}B.Nienhuis, {\sl J.Stat. Phys.}, {\bf 34}, 731 (1984)
\bibitem{Bernard82}B.Nienhuis, {\sl J.Phys.}, {\bf A15}, 199 (1982)
\bibitem{D76}J.Des Cloizeaux, {\sl J.de Phys.}, {\bf 37}, 431 (1976)
\bibitem{DG85}P.G.de Gennes, "Scaling Concepts in Polymer Physics", Cornell
University Press, Ithaca (1985)
\bibitem{Ber82}B.Duplantier, {\sl J.de Phys.}, {\bf 43}, 991 (1982)
\bibitem{B89}J.P.Bouchaud, A.Georges, {\sl Phys. Rev.}, {\bf B39}, 2846 (1989)
\bibitem{JB}J.B.Zuber, in "Perspectives in String Theory", P.di Vecchia et al.
Editors, World Scientific (1988)
\bibitem{MSS89}G.Mussardo, G.Sotkov, M.Stanishkov, {\sl Int. J. Mod. Phys.},
{\bf A4}, 1135 (1989)
\bibitem{DS87}B.Duplantier, H.Saleur, {\sl Nucl. Phys.}, {\bf B290}, 291 (1987)
\bibitem{WR89}N.Warner, in "Superstrings 89", Proceedings of Trieste Spring
School, M.Green et al. Editors, World Scientific (1989)
\bibitem{CK90}M.Cvetic, D.Kutasov, {\sl Phys. Lett.}, {\bf B240}, 61 (1990)
\bibitem{DS87bis}B.Duplantier, H.Saleur, {\sl Phys. Rev. Lett.}, {\bf 59}, 539
(1987)
\bibitem{FMV90}P.Fendley, S.D.Mathur, C.Vafa, N.Warner, {\sl Phys. Lett.}, {\bf
B243}, 257 (1990)
\bibitem{Zlast}Al.B.Zamolodchikov, "From Tricritical to Critical Ising by
Thermodynamic Bethe Ansatz", preprint ENS-LPS-327 (1991)
\bibitem{VR}R.Villanove, F.Rondelez, {\sl Phys. Rev. Lett.}, {\bf 45}, 1502
(1980)
\bibitem{V91}C.Vafa, {\sl Mod. Phys. Lett.}, {\bf A6}, 337 (1991)
\bibitem{CV91}S.Cecotti, C.Vafa, "Topological Antitopological Fusion", preprint
HUTP-91
\end{thebibliography}
\end{document}